%% file: main.tex
\documentclass[letterpaper,twocolumn,10pt]{article}
\usepackage{usenix,epsfig,endnotes}
\usepackage{booktabs}
\usepackage{makecell}
\usepackage{enumitem}
\usepackage{multirow}
\usepackage{xcolor}
\usepackage{tabularx}
\usepackage{booktabs}
\usepackage{xspace}
\begin{document}

\date{}

\definecolor{sageleaf}{RGB}{233, 237, 201}
\definecolor{blushcloud}{RGB}{241,231,231}
\definecolor{gainsboro}{RGB}{220,220,220}

\newcommand{\systemname}{\textsc{K-Repro}\xspace}

\title{\Large \bf Patch-to-PoC: A Systematic Study of Agentic LLM Systems for Linux Kernel N-Day Reproduction}
\author{
\begin{tabular}{c}
{\normalfont\rmfamily\upshape\large
Juefei Pu$^{*}$, Xingyu Li$^{*}$, Zhengchuan Liang$^{*}$, Jonathan Cox$^{*}$,
}\\
{\normalfont\rmfamily\upshape\large
Yifan Wu$^{*}$, Kareem Shehada$^{*}$, Arrdya Srivastav$^{*}$, Zhiyun Qian$^{*}$
}\\[0.35em]
{\normalfont\rmfamily\upshape\small $^{*}$University of California, Riverside}\\[0.2em]
{\normalfont\ttfamily\small
jpu007@ucr.edu, xli399@ucr.edu, zlian064@ucr.edu, jcox033@ucr.edu
}\\
{\normalfont\ttfamily\small
fshal003@ucr.edu, kareem.shehada@email.ucr.edu, asriv033@ucr.edu, zhiyunq@cs.ucr.edu
}
\end{tabular}
}

\maketitle

\thispagestyle{empty}

\begin{abstract}
Autonomous large language model (LLM) based systems have recently shown promising results across a range of cybersecurity tasks. However, there is no systematic study on their effectiveness in autonomously reproducing Linux kernel vulnerabilities with concrete proofs-of-concept (PoCs). Owing to the size, complexity, and low-level nature of the Linux kernel, such tasks are widely regarded as particularly challenging for current LLM-based approaches.

In this paper, we present the first large-scale study of LLM-based Linux kernel vulnerability reproduction. For this purpose, we develop \systemname, an LLM-based agentic system equipped with controlled code-browsing, virtual machine management, interaction, and debugging capabilities.  Using kernel security patches as input, \systemname automates end-to-end bug reproduction of N-day vulnerabilities in the Linux kernel. On a dataset of 100 real-world exploitable Linux kernel vulnerabilities collected from KernelCTF, our results show that \systemname can generate PoCs that reproduce over 50\% of the cases with practical time and monetary cost. 

Beyond aggregate success rates, we perform an extensive study of effectiveness, efficiency, stability, and impact factors to explain when agentic reproduction succeeds, where it fails, and which components drive performance. These findings provide actionable guidance for building more reliable autonomous security agents and for assessing real-world N-day risk from both offensive and defensive perspectives.

\end{abstract}

\input{sections/introduction.tex}
\input{sections/background.tex}

\input{sections/methodology.tex}

\input{sections/performancemeasurement.tex}

\input{sections/executioncontextmeasurement}

\input{sections/beatsyzdirect}

\input{sections/casestudies}

\input{sections/related}

\input{sections/conclusion}

\cleardoublepage
\bibliographystyle{plainurl}
\bibliography{sample}

\end{document}

%% file: sections/introduction.tex
\section{Introduction}

N-day vulnerability reproduction is an important task in cybersecurity. 
For attackers, vulnerability reproduction is a prerequisite for N-day exploit development, enabling them to exploit these vulnerabilities before corresponding patches are broadly deployed \cite{oh2009fight, wang2019detecting}. For defenders, it improves vulnerable-version identification \cite{tan2023syzdirect, dai2021facilitating, syzbridge} and supports remediation prioritization via exploitability assessment \cite{syzscope, jiang2023aem}.

To reproduce an N-day vulnerability, a common approach uses the security patch together with the target system as input and applies directed greybox fuzzing (DGF)\cite{bohme2017directed}. DGF leverages lightweight program instrumentation and distance metrics to steer input generation toward patch-related code regions. In the context of vulnerability reproduction, the code modified by the security patch is treated as the fuzzing target, while testing is performed on the corresponding unpatched version of the program, e.g., SyzDirect \cite{tan2023syzdirect} being the state-of-the-art for Linux kernel bug reproduction. 

Traditional fuzzing-based pipelines remain bottlenecked by harness and template engineering~\cite{winnie,difuze, syzdescribe, syzgen++, syzspec, dependency_challenge}: without high-quality, target-aware input descriptions, fuzzing effectiveness drops sharply. LLMs offer a complementary path to alleviate this bottleneck. Owing to their code-understanding and code-generation capabilities, recent work uses LLMs to automate parts of harness/template construction, and improve target reachability in directed fuzzing \cite{kernelgpt, yourexp, kim2025atlantis, sheng2025all}, or even directly generate working PoCs~\cite{zeng2025pbfuzz,simsek2025pocgen,nitin2025faultline,ullah2025cve,CyberGym}.
Beyond academic prototypes, DARPA's AI Cyber Challenge (AIxCC) \cite{aicyberchallengeAicyberchallengecom} aims to leverage LLM-powered systems to derive PoCs from bug-introducing commits (instead of from patches). The success of AIxCC demonstrates that with the power of LLMs, cyber-reasoning systems can outperform traditional methods and identify and reproduce vulnerabilities in large codebases. 

However, despite LLM-based approaches having achieved notable success, most of these works are for user-space programs, and very limited studies have been applied to the Linux kernel domain. ReachabilityAgent \cite{reachabilityagent} leverages LLM to generate inputs that exercise target functions, which is a premise for vulnerability reproduction, but it only succeeded in one out of five cases; the limited success is explained as ``due to the inherent complexity of the target code''. In the AIxCC competition, only a single team solved the Linux kernel challenge in the semi-final competition \cite{whexyLostAIxCC}, and the kernel challenges were excluded from the final competition. Thus, it appears that LLMs are not yet powerful enough to handle the Linux kernel and perform vulnerability reproduction. To date, SyzDirect \cite{tan2023syzdirect} remains the state-of-the-art solution for this task, and it relies solely on DGF rather than LLMs.

In this paper, we provide the first systematic large-scale evaluation of LLM-agent-based vulnerability reproduction in the Linux kernel.
To this end, we develop \systemname, an LLM-based agentic system equipped with a small but expressive toolset, including code browsing, virtual machine management, virtual machine interaction, and GDB debugging.
Our results on a large real-world dataset of 100 KernelCTF vulnerabilities~\cite{kernelctf} show that LLMs are, in fact, capable of performing kernel bug reproduction. Specifically, our results show consistently higher than 50\% success rates under various setups, even for bugs that are reported after the knowledge cutoff date of the LLM models we used.
We find that LLM agents are good at invoking shell commands and library APIs to manipulate the kernel subsystem in which the bug lives. They are also generally good at looking for feedback and adjusting the PoCs iteratively.
Furthermore, \systemname achieves significantly higher effectiveness and substantially faster reproduction 
than the state-of-the-art directed greybox fuzzing approach, SyzDirect~\cite{tan2023syzdirect}. Beyond aggregate performance, we perform in-depth analyses of both successful and failed cases to identify key contributing factors and limitations, providing concrete insights into the practical strengths and weaknesses of LLM-based approaches for kernel bug reproduction.

Our contributions are summarized as follows:

\begin{itemize} [nosep]
  \item We design and implement \systemname, a fully automated, reusable, and extensible LLM-based agentic system for Linux kernel vulnerability reproduction. \systemname takes as input only a patch (e.g., a bug-fixing commit). It then builds a vulnerable kernel in a VM and generates a PoC without querying the Internet. We will open-source \systemname to facilitate future research.

  \item Using \systemname, we conduct the first large-scale empirical study of LLM-based Linux kernel vulnerability reproduction on real-world vulnerabilities, achieving over 50\% success rate in the 100-bug KernelCTF dataset and 64\% in repeated runs. We measure effectiveness, efficiency, and stability to provide a quantitative assessment, and show that our approach substantially outperforms fuzzing-based approaches.

  \item We systematically stress-test \systemname under capability restrictions (e.g., no debugging) and degraded input signals, and quantify their causal impact on reproduction trajectories and final success, yielding actionable design principles and a strong empirical foundation for next-generation kernel vulnerability reproduction agents.

\end{itemize}

%% file: sections/background.tex
\section{Background}
\subsection{Linux Ecosystem}

The Linux kernel is the foundation of much of today's computing, from cloud servers to mobile devices. 
As a common infrastructure, it implements a broad set of features across many subsystems (e.g., filesystems, networking, and device drivers).
This creates a large attack surface, which in turn leads to a steady stream of discovered vulnerabilities. 
Its monolithic design means a single kernel bug can compromise the whole system. 

The situation is further exacerbated by the Linux kernel ecosystem~\cite{li2024investigation}. In this ecosystem, Linux mainline serves as the upstream development branch, stable/LTS branches~\cite{kernelEverythingEver} are forked from mainline, and downstream Linux distributions (e.g., Ubuntu, Android) further fork these stable/LTS branches and maintain their own kernels for end users. Consequently, when a security fix lands in the upstream kernel, it takes time to be ported across multiple stable/LTS branches and then into downstream kernels. It often takes dozens of days for security patches to be ported~\cite{li2024investigation,zhang2021investigation}, and in some cases they remain unported for more than half a year due to being overlooked~\cite{CVE202352620}.

Such a time gap opens up a window for attackers to exploit N-day vulnerabilities against downstream systems  \cite{projectzeroMoreKnow}, when patches are already available in upstream (and thus available to attackers for analysis and exploit development) but not yet available in downstream kernels \cite{maar2025doom,woo2023v1scan}. Indeed, such attacks have occurred in the wild, e.g., \cite{projectzeroAnalyzingModern}.

\subsection{PoC Generation for Linux Kernel}

Automated PoC generation can accelerate exploit development for N-day vulnerabilities. At the same time, it helps defenders triage vulnerabilities and prioritize patches.
PoC generation for the Linux kernel is significantly different from and considerably more challenging than that for user-space programs. First, the Linux kernel is extremely large and exposes highly heterogeneous and subsystem-specific interfaces. As a result, effective kernel fuzzing for a target subsystem typically requires carefully tailored syscall descriptions \cite{syzkaller,lkl}. Although a number of solutions aim to automate the syscall description generation \cite{difuze, syzdescribe, syzspec, syzgen++, kernelgpt}, their completeness and correctness are still not guaranteed. Moreover, the Linux kernel operates in a highly stateful and multi-threaded manner, which causes most bugs to require multiple syscalls to trigger \cite{dependency_challenge,zhang2025statically} and often involve race conditions.
This significantly increases the difficulty of crash reproduction and PoC generation for fuzzers \cite{yome}. 

Although LLMs have shown effectiveness in user-space PoC generation and kernel fuzzing template generation \cite{kernelgpt}, producing kernel inputs from scratch \cite{reachabilityagent} is still regarded as challenging for LLMs. To date, we have not seen LLMs being effective at vulnerability reproduction, where PoCs must be precisely targeted to specific kernel code regions.

%% file: sections/methodology.tex
\section{Methodology}

In this section, we introduce the goals of our study, the \systemname system we developed to conduct the study, and the generic setups.

\subsection{Goals}
LLM agents, when equipped with appropriate tools and instructions, are increasingly capable of performing complex software engineering tasks. In this work, our goal is to systematically understand and characterize the capability of LLM agents in reproducing Linux kernel vulnerabilities.
Specifically, we are interested in answering three research questions:

\textbf{RQ1:} What is the end-to-end characterization of LLM-based Linux kernel vulnerability reproduction?

\textbf{RQ2:} How do restrictions of toolset and input signals affect agents' performance in Linux kernel vulnerability reproduction?

\textbf{RQ3:} How does the performance of LLM-based vulnerability reproduction compare with state-of-the-art non-LLM approaches such as SyzDirect?

To answer these research questions, we design and conduct three corresponding sets of studies using the agentic system we developed, i.e., \systemname.

\subsection{\systemname Architecture}

\begin{figure}[htbp]
  \centering
  \includegraphics[width=\linewidth,trim=10 16 10 18, clip]{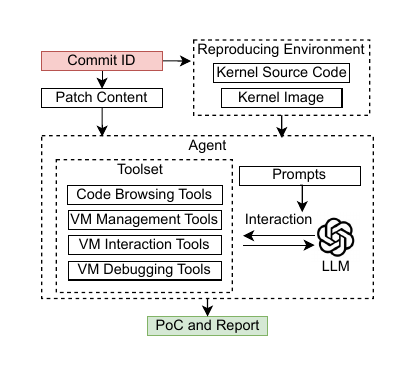}
  \caption{\label{fig:arch}Overall Architecture }
\end{figure}

The overall architecture of \systemname is depicted in Figure~\ref{fig:arch}. To support multiple sets of experiments, \systemname is designed with a minimal and standardized interface. As a result, the only external input to \systemname is a security patch commit identifier, based on which it automatically performs environment deployment and vulnerability reproduction, and outputs the generated PoC, analysis report, and execution logs.

Given a security patch commit, \systemname first extracts the patch content and prepares a reproducing environment. This environment includes the kernel source code and a runnable kernel image that is \emph{one commit prior to the security patch}, which serves as the execution context for subsequent analysis and PoC validation. The preparation of the reproducing environment is performed using a deterministic, non-LLM pipeline, ensuring reproducibility and stability across runs.

The core of \systemname is an autonomous agent that couples an LLM with a small but powerful toolset, enabling the LLM to perform vulnerability reproduction in an agentic manner. Under the instruction of prompts, the agent interacts with the target kernel through tool invocations to perform the vulnerability reproduction task. The agent iteratively performs static code analysis through code browsing tools, generates PoCs with the help of shell command tools, obtains dynamic feedback and validates PoCs via virtual machine (VM) interaction tools, and inspects kernel states through VM debugging tools. 

After task completion, \systemname retains not only the generated PoC and analysis report, but also the raw execution trace that records all tool invocations and virtual machine logs. These artifacts provide a detailed and auditable record of the vulnerability reproduction process and are essential for subsequent analysis.

\subsection{\systemname Toolset}
We provide \systemname with four types of tools listed below, all of which are commonly leveraged by human vulnerability researchers. 
\noindent\textbf{Code Browsing Tools:} Code browsing tools allow the agent to inspect kernel source code in an on-demand manner. Using these tools, the agent can list symbols defined in large source files, globally query definitions and references of specific symbols, and retrieve code snippets from designated files and line ranges. This design enables the agent to acquire sufficient contextual information about the vulnerability while avoiding unnecessary exposure to large volumes of source code that could exhaust the context window.

\noindent\textbf{VM Management Tools:}
VM management tools support lifecycle management for the virtual machine used in dynamic testing. Since the virtual machine may become unresponsive during testing, these tools expose explicit operations for starting and restarting the VM, allowing the agent to recover from errors automatically. VM startup is snapshot-based, ensuring that each execution begins from an identical memory state and reducing nondeterminism across runs. In addition, \systemname provides a utility that compiles C source code and uploads the resulting binary to a designated directory inside the VM in a single step, integrating the PoC generation and uploading in a streamlined process.

\noindent\textbf{VM Interaction Tools:}
VM interaction tools allow the agent to directly interact with the running VM. The agent can send commands to and receive outputs from the VM console, as well as issue control signals (e.g., interrupt or break) to terminate programs running inside the guest. Kernel messages and console outputs are multiplexed into the same interaction channel, allowing the agent to observe kernel outputs alongside command results without invoking additional tools.

\noindent\textbf{VM Debugging Tools:} VM debugging tools expose low-level kernel debugging capabilities to the agent. Through these tools, the agent can inspect kernel registers and memory, set and manage breakpoints, and execute raw GDB commands against the running kernel. Notably, debugging tools can detect when the kernel hits a breakpoint and suspend the agent's access to VM interaction tools. This ensures that the agent correctly attributes the halted state to the breakpoint rather than as a kernel hang.

\subsection{\systemname Prompts}
Beyond specifying the task (i.e., producing a working PoC and a reproduction report), \systemname uses prompts that provide both high-level and technical guidance.

\noindent\textbf{High-level Guidance:} This part specifies the agent's behaviors in a high-level way. The LLM is instructed to perform static analysis before dynamic validation, to form hypotheses and revise or reject them based on evidence, and to remain persistent until it reproduces the vulnerability. Although these instructions are high-level, they may still affect the trajectory of LLM and influence the overall effectiveness. We evaluate them in Section~\ref{sec:rq23-badprompt}.

\noindent\textbf{Technical Guidance:} This part provides operational instructions for interacting with \systemname's tools. Some rules improve efficiency, such as recommending setting shorter timeouts and discouraging expensive actions (e.g., full kernel memory searches). Other rules mitigate recurring misuse patterns observed during development. For instance, the prompts prohibit the agent from accessing the Internet to search for PoCs. Although \systemname has built-in control on Internet access, explicitly stating these constraints helps the agent avoid making attempts and remain focused on the intended workflow.  

\subsection{\systemname Implementation}
\systemname is implemented using FastMCP\cite{fastmcp}, with all tools integrated into a unified Model Context Protocol (MCP) server. The reproducing environment is built on QEMU, kernel symbol lookup and reference queries are supported via CodeQuery\cite{codequery}, and kernel debugging is implemented through GDB MI\cite{gdbmi}. The agent client is based on the Codex CLI, which has a built-in tool to execute host-side shell commands. We exclude host-side shell commands from our statistical analysis because we did not observe a significant impact on the results. 
The agent runs with Codex's built-in sandbox turned off in a containerized environment, ensuring both flexibility and safety.

\subsection{Basic \systemname Setup}
\label{sec:measurementsettings}

For each vulnerability we test, the vulnerable kernel image is built using GCC~10 with the kernel configuration adopted by the state-of-the-art kernel fuzzing cluster Syzbot \cite{syzbot}. 
This configuration provides broad compatibility across kernel subsystems and enables KASAN, allowing memory corruption to be detected immediately when triggered. 

Each run in \systemname is allocated a maximum time budget of 10 hours with no monetary cost limit. \systemname uses the state-of-the-art reasoning model GPT-5.1 Codex Max with XHigh reasoning level as the primary model, and GPT-5.1 Codex with Medium reasoning level for comparison. The knowledge cutoff date for both models are Sep 30, 2024. For simplicity, we refer to these two models as XHigh and Medium in the remainder of this paper.

To support large-scale evaluation, reproduction success is determined using an automated analysis script that checks for the presence of a PoC and report, and parses virtual machine logs for kernel crashes. All experiments are conducted purely from user space without loading additional kernel modules. Under this setting, a kernel-level crash indicates that a kernel vulnerability has been triggered.

Due to the multi-behavior nature of Linux kernel vulnerabilities~\cite{lin2022grebe,syzscope}, it is difficult for automated analysis to verify root-cause equivalence between crashes. Therefore, following common practice, any kernel-level crash is treated as a successful reproduction.

We therefore manually inspect kernel logs and tool invocation trajectory for all successful case with XHigh in Section~\ref{sec:tool-usage} to confirm that the crashes occur at the same vulnerability point, and explicitly analyze cases involving ``cheating'' as illustrated in Section~\ref{sec:tool-usage}. Such cases are marked as failures and we encode rules of detecting this in our automated result analysis script for all other runs.

%% file: sections/performancemeasurement.tex
\section{Reproduction Performance}
\label{sec:performeasure}

To answer \textbf{RQ1} and characterize the agent's performance in Linux kernel vulnerability reproduction, we conduct a set of performance measurements along the following dimensions: 

\textbf{RQ1.1:} How well does agent perform in real-world Linux kernel vulnerability reproduction in terms of effectiveness, efficiency, and monetary cost?

\textbf{RQ1.2:} How different tools are used in the reproduction?

\textbf{RQ1.3:} What properties of Linux kernel vulnerabilities affect the likelihood of successful reproduction? 

\textbf{RQ1.4:} How robust and convergent is the agent performance under repeated runs?

\subsection{Experiment Setup}
\label{sec:expsetoverall}

\paragraph{Dataset}
We construct our evaluation dataset from cases in Google’s KernelCTF~\cite{kernelctf}, a vulnerability rewards program (VRP) that invites security researchers to demonstrate exploitation techniques for Linux kernel 0-day and 1-day vulnerabilities across multiple target instances.

We choose KernelCTF as the evaluation dataset for three reasons. 
First, each KernelCTF bug corresponds to a real-world exploitable Linux kernel vulnerability, which aligns with our goal of measuring vulnerability reproduction capability, whereas general Linux CVE datasets include potential bugs rather than these proved to be triggerable \cite{kernelCVEsx2014}. 
Second, KernelCTF provides public writeups and exploit source code for many cases after  grace periods of fixing, forming ground truths and facilitating manual analysis of the agent's behavior.
Third, KernelCTF cases span vulnerabilities disclosed between 2023 and 2025, allowing us to study the impact of LLM knowledge cutoff dates. 

We include all KernelCTF cases released before November~24,~2025, and retain only those whose fixing commits are available in the Linux upstream repository and whose vulnerable kernels can boot successfully in our virtual machine setup environment. This results in 100 evaluation cases out of the original 116 cases.

\paragraph{Initial Evaluation.} We perform the initial evaluation on KernelCTF bugs. For each KernelCTF bug, we feed the corresponding upstream kernel commit to the \systemname pipeline and perform one run of XHigh and Medium. Data collected here enables us to analyze overall performance, tool usage, performance by property. 
\paragraph{Stability and Convergence Evaluation.} 
To analyze performance stability, we repeat the experiment with XHigh for all cases in a second run under identical measurement settings, examining the reproducibility and stability.  
To evaluate the convergence of the aggregated success rate, we identify cases that fail in both runs and re-execute only these remaining failed cases in an iterative manner. This process is repeated until no additional successful reproductions are observed.

\subsection{Success Rates and Costs}

To answer \textbf{RQ1.1}, we analyze the success rate, execution time per case, and monetary cost per case based on a single execution of the dataset for each model.

We begin with an aggregate characterization of overall performance. Table~\ref{tab:overall-stats} summarizes the results for each model by grouping evaluation cases into successful and failed runs, as well as all cases combined. For each group, we report the average execution time and monetary cost per case, which together provide a coarse-grained view of effectiveness and efficiency before more fine-grained analysis.

\begin{table}[htbp]
\centering
\caption{Overall performance statistics on time and cost }
\label{tab:overall-stats}
\scalebox{0.9}{
\begin{tabular}{c c c c c}
\toprule
Model & Result & counts & \makecell{Avg. Cost \\ (USD)} & \makecell{Avg. Time \\ (min)} \\
\midrule
Medium & all & 100 & 4.66 & 33.80 \\
Medium & success & 47 & 2.57 & 18.67 \\
Medium & fail & 53 & 6.52 & 47.22 \\
\midrule
XHigh  & all     & 100 & 4.02 & 19.33 \\
XHigh  & success & 56  & 2.49 & 11.89 \\
XHigh  & fail    & 44  & 5.96 & 28.81 \\
\bottomrule
\end{tabular}
}
\end{table}

Across the 100 KernelCTF cases, the XHigh model successfully reproduces 56.0\% of vulnerabilities, while the Medium model succeeds in 47.0\% of cases. These results indicate that, under our measurement settings, agents are able to reproduce a substantial fraction of real-world exploitable Linux kernel vulnerabilities without human intervention.

Including failed attempts, the average monetary cost per case is \$4.7 for Medium and \$4.0 for XHigh. In terms of execution time, Medium requires an average of 33.8 minutes per case, while XHigh completes a case in 19.3 minutes on average.

However, aggregate averages obscure substantial variability across cases. For Medium, the maximum monetary cost reaches \$14.3 in successful cases and \$31.3 in failed cases, with corresponding runtimes of up to 127.6 and 299.7 minutes. By contrast, XHigh exhibits lower maxima, with costs of \$8.0 and \$10.9 and runtimes of 33.9 and 70.7 minutes for successful and failed cases, respectively.

\begin{figure}[htbp]
  \centering
  \includegraphics[width=0.9\linewidth]{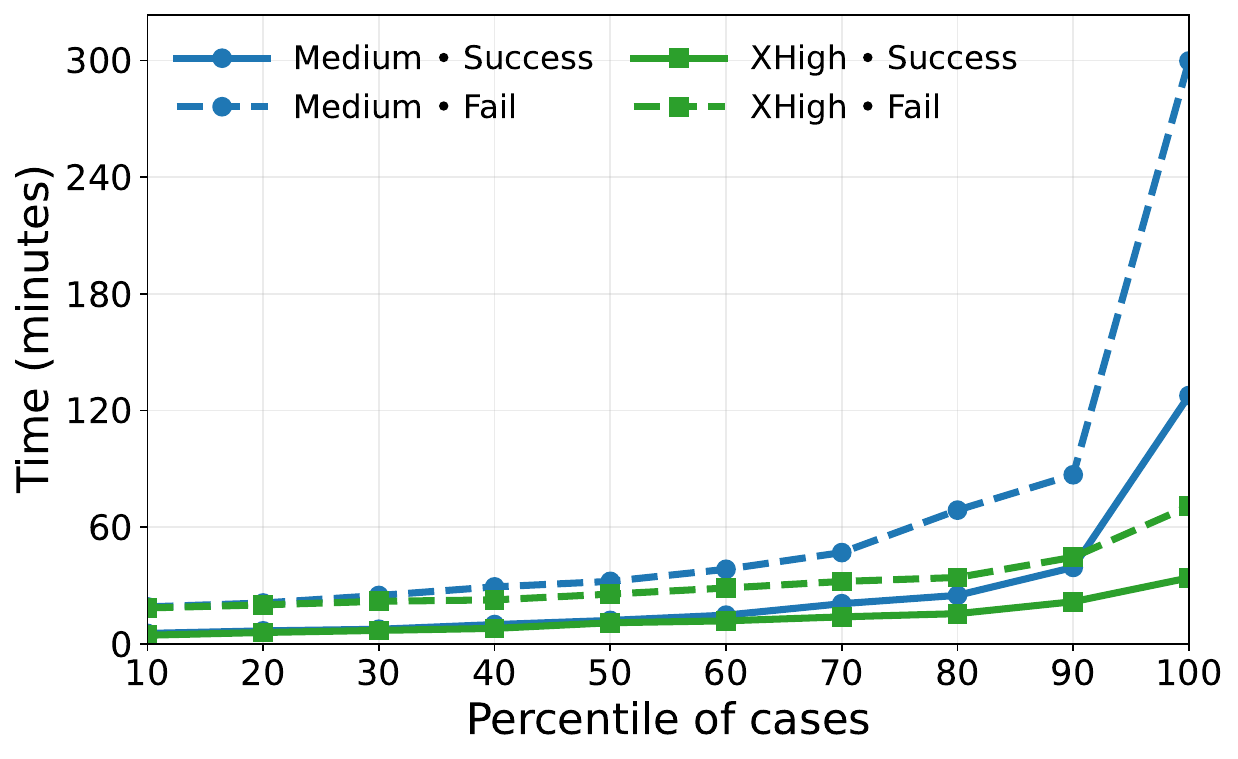}
  \caption{\label{fig:timepercent}Time distribution across case percentiles.}
\end{figure}

As shown in Figure~\ref{fig:timepercent}, successful reproductions typically complete within tens of minutes. For Medium, 90\% of successful cases finish within 39.2 minutes, with the slowest success taking 127.6 minutes; for XHigh, 90\% of successful cases finish within 21.6 minutes, and all successful cases complete within 33.9 minutes. Consequently, a one-hour per-case time budget would preserve all successful XHigh runs and exclude only two successful Medium cases.

In contrast, failed runs exhibit markedly different runtime behavior: Medium shows heavy-tailed failures with some executions extending to nearly five hours, whereas XHigh maintains a much tighter runtime distribution without comparable long-running failures.

Overall, our measurements for \textbf{RQ1.1} show that LLM-based vulnerability reproduction achieves non-trivial success rates with per-case runtimes on the order of tens of minutes and monetary costs of only a few dollars on average. This demonstrates the effectiveness and efficiency of agents for real-world vulnerability reproduction. 

\subsection{Tool Usage}
\label{sec:tool-usage}

To answer \textbf{RQ1.2} we analyze the tool usage in both runs.

\paragraph{Overall stats.} We analyze how \systemname uses external tools. On average, Medium executes 165.1 tool calls per case, including 11.7 VM management actions and 91.7 VM interactions, 38.7 source-browsing actions, 23.0 debugging actions; XHigh executes 129.5 tool calls per case, including 9.2 VM management actions and 55.5 VM interactions, 40.5 source-browsing actions, 24.4 debugging actions. This result shows that static analysis, dynamic verification, and debugging are all well leveraged. 

Next, to further understand the agent behaviors, we break down the usage patterns for each tool category, highlighting notable phenomena in the following paragraphs.

\paragraph{PoC Iterations.} We analyze the number of PoC generation and uploading attempts required by agents in each case.

On average, XHigh attempted to generate PoCs with an average of 7.4 iterations per case, compared to 9.1 for Medium. Successful cases require substantially fewer iterations on average, with 4.9 for XHigh and 4.5 for Medium. 
Interestingly, a non-trivial fraction of successful reproductions complete with one single PoC compilation and upload: 10.7\% for XHigh and 21.3\% for Medium.
Among one-shot successes, 3/6 for XHigh and 3/10 for Medium correspond to patches submitted after the knowledge cutoff date of the models, suggesting that these successes are unlikely to be purely due to memorization of known PoCs. The effects of knowledge cutoff date is further studied in Section~\ref{sec:knowledge-cutoff}.

These results indicate that one-shot reproduction is uncommon, and agents typically rely on progressively refining PoCs to achieve successful vulnerability reproduction.

\paragraph{GDB Usage.}
As expected, we observe that GDB is commonly used by both models. XHigh sets breakpoints in 62 out of 100 cases, while Medium does so in 60 out of 100 cases. In most of these runs, breakpoints are successfully hit at least once: 59 of the 62 cases for XHigh and 54 of the 60 cases for Medium. Notably, this hit rate does not decrease in failed reproductions, where all 29 failed XHigh cases and 18 of the 22 failed Medium cases with breakpoints successfully trigger them. This indicates that reachability is usually not a bottleneck for kernel vulnerability reproduction.

Across all runs, stepping is almost never used, with only three stepping commands observed in total. In contrast, variable inspection is common, with an average of 5.4 inspections per case for XHigh and 4.4 for Medium. This suggests that agents prefer to inspect key program state at selected execution points rather than performing instruction-level stepping.

We also observe four cases where agents attempt to directly modify execution behavior through GDB, such as altering kernel memory or invoking kernel functions, resulting in artificial (“fake”) crashes.
We classify such cases as cheating and treat them as failures.

\paragraph{Command-Line Invocations.}
Beyond executing the generated PoCs, the agent also performs other interactions via the VM through command lines to facilitate the reproduction process. 

We find that the agent frequently invokes subsystem-specific command-line utilities. In particular, \texttt{nft} and \texttt{tc} are heavily used in both settings, with an average of 4.5 and 3.1 invocations per case for XHigh, and 5.3 and 4.1 invocations per case for Medium, respectively. This is likely attributable to the fact that these tools provide direct command-line access to kernel subsystems (e.g., Netfilter and \texttt{net/sched}), which lets the agent configure kernel state and observe higher-level, semantically meaningful outcomes.

We also quantify how often agents probe the kernel-exposed state through \texttt{procfs} and \texttt{sysfs}, some of which are also tied to specific subsystems. In XHigh, 31 cases use them (5.6 accesses on average among those runs), while Medium utilizes them in 33 runs with higher intensity (7.2 accesses on average). Aggregating accesses reveals two distinct purposes. Beyond checking kernel configuration and symbol availability (e.g., via \texttt{/proc/config.gz} and \texttt{/proc/kallsyms}), we observe that the agent notably accesses runtime kernel state through \texttt{procfs}. In particular, \texttt{/proc/crypto} is accessed 26 times in Medium and 49 times in XHigh, and \texttt{/proc/net} is accessed 13 times in Medium. This shows that agents would leverage them to analyze the state of the crypto API and network-related subsystems.

Overall, our analysis for \textbf{RQ1.2} shows that the agent performs vulnerability reproduction through multiple iterative attempts that combine static analysis and dynamic validation, actively leverages command-line utilities to interact with kernel subsystems, and primarily uses GDB to inspect key execution states rather than for instruction-level stepping.

\subsection{Vulnerability-Specific Factors Analysis}
\label{sec:vulfact}

To answer \textbf{RQ1.3}, we analyze how different properties of a vulnerability correlate with the LLM's likelihood of achieving successful vulnerability reproduction. Specifically, we examine several vulnerability-specific factors, including the affected kernel subsystem, vulnerability type, the presence of race conditions.

\paragraph{Subsystem-Specific Performance.}
We first analyze subsystem-specific performance to examine whether \systemname performs differently across kernel subsystems,
which accept different input formats and have varying degrees of complexity. For example, some accept \texttt{ioctl} commands with relatively simple arguments while others accept complex nested \texttt{netlink} messages, which are hard to construct~\cite{nlsaber}. Meanwhile, the LLM's knowledge about these subsystems may also vary, depending on the availability of documentation. Therefore, we group cases by kernel subsystem and conduct statistical analysis.  We leverage the odds ratio (OR)

to quantify the effect of subsystems on success probability and apply Fisher’s exact test\cite{fisher1922interpretation} to assess the statistical significance of the observed associations. The result is summarized in Table~\ref{tab:rq13-subsystem-combined}.

\begin{table}[htbp]
\centering
\small
\setlength{\tabcolsep}{4pt}
\begin{tabular}{llcc}
\toprule
Subsystem & Model & Success Rate & OR (p) \\
\midrule
net & Medium & 24/50 (48.0\%) & 1.08 (1.000) \\
    & XHigh  & 28/50 (56.0\%) & 1.00 (1.000) \\
\midrule
netfilter & Medium & 14/34 (41.2\%) & 0.70 (0.526) \\
          & XHigh  & 18/34 (52.9\%) & 0.83 (0.676) \\
\midrule
bpf & Medium & 4/7 (57.1\%) & 1.55 (0.703) \\
    & XHigh  & 4/7 (57.1\%) & 1.05 (1.000) \\
\midrule
others & Medium & 5/9 (55.6\%) & 1.46 (0.731) \\
       & XHigh  & 6/9 (66.7\%) & 1.64 (0.727) \\
\bottomrule
\end{tabular}
\caption{Statistics on subsystem impact on success rate (Fisher exact OR vs. remaining subsystems, two-sided p)}
\label{tab:rq13-subsystem-combined}
\end{table}

According to the results, neither model exhibits a substantial performance degradation across different subsystems. 
Although the success rates on netfilter are lower for both models (41.2\% for Medium and 52.9\% for XHigh), 
the corresponding p-values (0.53 for Medium and 0.68 for XHigh) indicate that these differences are not statistically significant.  Overall, the results suggest that \systemname does not exhibit a clear subsystem-specific weakness across the Linux kernel. One possible explanation is that all these subsystems have been heavily targeted in the wild, with sufficient publicly available documentation on exploitation techniques and kernel subsystem analysis.

\paragraph{Race-Condition Vulnerabilities.}

Reproducing race-condition vulnerabilities is generally considered more challenging, as it requires reasoning about concurrent executions and coping with the inherent nondeterminism of dynamic execution. To assess whether and how such characteristics impact LLM-based vulnerability reproduction, we compare success rates between race-condition and non-race cases in Table~\ref{tab:rq13-race-rates}.

\begin{table}[htbp]
\centering
\small
\setlength{\tabcolsep}{4pt}
\begin{tabular}{llc}
\toprule
Vulnerability Type & Model & Success Rate \\
\midrule
Race & Medium & 5/24 (20.8\%) \\
     & XHigh  & 7/24 (29.2\%) \\
\midrule
Non-race & Medium & 42/76 (55.3\%) \\
         & XHigh  & 49/76 (64.5\%) \\
\bottomrule
\end{tabular}
\caption{Statistics on concurrency’s impact on success rate.}
\label{tab:rq13-race-rates}
\end{table}

According to the table, race-condition vulnerabilities exhibit substantially lower success rates for both models.
For the Medium model, race-condition cases achieve a success rate of 20.8\%, compared to 55.3\% for non-race cases.
A similar gap is observed for XHigh, where the success rate is 29.2\% for race-condition cases versus 64.5\% for non-race cases. 

As expected, race-condition vulnerabilities are significantly more difficult to reproduce than non-race ones. Future approaches may benefit from combining LLM-based methods with complementary techniques, such as directed fuzzing, to better handle such cases.

\paragraph{Vulnerability Type.}
We next examine whether vulnerability type affects the likelihood of successful reproduction.
From a root-cause perspective, memory corruption vulnerabilities are broadly categorized into \emph{temporal} and \emph{spatial} memory violations\cite{huang2024top}.
Temporal memory violations, exemplified by use-after-free (UAF) and double-free (DF) vulnerabilities,
require reasoning about object lifetime and variable liveness across different execution contexts,
whereas spatial memory violations, represented by out-of-bounds (OOB) accesses,
primarily involve tracking the values and bounds of specific memory objects. Such a distinction has been shown to make temporal memory vulnerabilities more challenging to analyze in practice \cite{lee2015preventing}.
We therefore hypothesize that this additional analytical complexity may also negatively impact the effectiveness of LLM-based vulnerability reproduction.

Motivated by this distinction, we group vulnerabilities into two categories, OOB and UAF/DF,
and focus our analysis on these dominant types, excluding the remaining two rare categories
that account for only 2 out of 100 cases.

A key confounder is that race conditions are not uniformly distributed across types:
among the retained 98 cases, 26.8\% of UAF/DF cases are race-condition vulnerabilities (22/82),
whereas only 6.3\% of OOB cases are race-condition vulnerabilities (1/16).
To disentangle the vulnerability-type effect from this correlation, we stratify cases by whether they are
race-condition vulnerabilities and report the stratum-specific counts in Table~\ref{tab:rq13-bugtype-stratified}.

\begin{table}[htbp]
\centering
\small
\setlength{\tabcolsep}{4pt}
\begin{tabular}{llcc}
\toprule
Race Stratum & Model & OOB (S/F) & UAF/DF (S/F) \\
\midrule
Race     & Medium & 1/0   & 4/18  \\
         & XHigh  & 1/0   & 6/16  \\
\midrule
Non-race & Medium & 13/2  & 29/31 \\
         & XHigh  & 13/2  & 36/24 \\
\bottomrule
\end{tabular}
\caption{Stratified success/failure (S/F) counts by vulnerability type under race and non-race settings (excluding rare types).}
\label{tab:rq13-bugtype-stratified}
\end{table}

We then aggregate the stratum-specific associations using the Mantel--Haenszel estimator and apply the
Cochran--Mantel--Haenszel (CMH) test for significance.
After controlling for race conditions, UAF/DF remains significantly harder than OOB for both models:
the race-adjusted odds ratios (UAF/DF vs.\ OOB) are 0.13 ($p = 0.002$) for Medium
and 0.20 ($p = 0.023$) for XHigh. This proves that even excluding the factor of race condition, temporal memory bugs are statistically significantly more challenging for agents to reproduce.

This observation is consistent with the intuition that temporal memory violations require reasoning about object lifetime
and temporal safety across extended execution contexts, which may be more demanding for current LLM-based approaches.

\paragraph{Commit Messages.}
Finally, we examine how the information disclosed in commit messages affects reproduction outcomes.
We categorize commit messages into three levels:

\begin{enumerate}[nosep]
  \item Messages that do not acknowledge a triggerable issue, such as framing the commit as simplification of code or describing changes without explaining reasons.
  \item Messages that acknowledge the existence of a triggerable issue but do not describe how to reproduce it.
  \item Messages that explicitly acknowledge the issue and provide concrete triggering steps.
\end{enumerate}

We manually audit commit messages and assign each case to one of the three levels,
aggregating the labels and reproduction outcomes in Table~\ref{tab:commitmsg-level-success}. 

\begin{table}[ht]
\centering
\small
\caption{Reproduction success rates under different commit-message disclosure levels.}
\label{tab:commitmsg-level-success}
\begin{tabular}{lccc}
\toprule
 Model & Type~1 & Type~2 & Type~3 \\
\midrule
XHigh  & 6/13 (46.2\%) & 44/79 (55.7\%) & 6/8 (75.0\%) \\
Medium & 3/13 (23.1\%) & 38/79 (48.1\%) & 6/8 (75.0\%) \\
\bottomrule
\end{tabular}
\end{table}

According to the results, richer information disclosure is consistently associated with higher reproduction success. Notably, the majority of cases fall into Type~2, where messages acknowledge a triggerable issue but do not provide concrete steps and agents still achieve strong success rates in the category. This suggests that reproduction success does not solely rely on directly replaying the explicit reproducing instructions from commit messages. We further analyze the impact of removing these commit messages in Section~\ref{sec:rq21-commitmsg}.

\subsection{Knowledge Cutoff Effects}
\label{sec:knowledge-cutoff}
Even though we do not allow \systemname to access the Internet during PoC generation, there may still be ``data leakage'' given that many KernelCTF bugs do come with the corresponding technical writeups and exploit source code before the knowledge cutoff date.
It is thus crucial to understand whether an agent's apparent effectiveness primarily stems from LLM being memorizing publicly available exploit details.

We can measure this effect by observing if noticeably higher
success rates on vulnerabilities disclosed before its knowledge cutoff (Sep 30, 2024), and degraded performance on those disclosed afterwards.
We approximate the disclosure timeline using the \emph{patch commit submission time} recorded in the KernelCTF reports,
and split cases into pre-cutoff (58 cases) and post-cutoff (42 cases).

Before comparing success rates, we verify that major confounders are not unevenly distributed across the two groups.
Race-condition cases are well-balanced (24.1\% pre-cutoff vs.\ 23.8\% post-cutoff; Fisher's exact test OR=0.98, $p=1.00$),
and the distribution of major vulnerability types is also similar (OOB: 15.5\% pre-cutoff vs.\ 17.5\% post-cutoff;
OR=1.15, $p=0.789$), mitigating concerns that one side is systematically harder due to case composition.

\begin{table}[htbp]
\centering
\small
\setlength{\tabcolsep}{4pt}
\begin{tabular}{llc}
\toprule
Cutoff Group & Model & Success Rate \\
\midrule
Pre-cutoff  & Medium & 28/58 (48.3\%) \\
            & XHigh  & 34/58 (58.6\%) \\
\midrule
Post-cutoff & Medium & 19/42 (45.2\%) \\
            & XHigh  & 22/42 (52.4\%) \\
\bottomrule
\end{tabular}
\caption{Success rates before vs.\ after the knowledge cutoff date (Sep 30, 2024).}
\label{tab:rq13-cutoff-rates}
\end{table}

We then apply Fisher's exact test (two-sided) to compare pre- vs.\ post-cutoff success rates for each model, and the result is summarized in Table~\ref{tab:rq13-cutoff-rates}. According to the table, the success rate for Medium decreased from 48.3\% to 45.2\%, while for XHigh decreased from 58.6\% to 52.4\%. Calculating Fisher's exact test, this result is not statistically significant for either Medium (OR = 1.13, $p = 0.84$) or XHigh (OR = 1.29, $p = 0.55$).

To summarize, our results for \textbf{RQ1.3} show that reproduction success is not sensitive to kernel subsystems and knowledge cutoff dates, but is significantly reduced for race-condition and temporal memory vulnerabilities (UAF/DF), and availability of informative commit messages.

\subsection{Stability and Convergence Analysis}

To answer \textbf{RQ1.4}, we repeat the measurement using XHigh under identical settings and compare the aggregate results with the initial run.

The average execution time varies by about 18\% across runs (19.33 $\rightarrow$ 22.85 ), while the average monetary cost differs by less than 5\%. The success rate decreases slightly from 56.0\% to 52.0\%. Across two runs, 11 cases succeed only in the initial run, while 7 cases succeed only in the repeated run. Taken together, 63 out of 100 cases succeed in at least one run, significantly exceeding the per-run success rates.

To evaluate convergence, we iteratively re-execute only the cases that fail in both runs.
The third execution yields one additional successful reproduction, while further repetitions produce no new successes. Thus the aggregated success rate for repeated runs is 64\%.

Overall, our evaluation for \textbf{RQ1.4} indicate that while aggregate success rates are relatively stable, individual vulnerabilities may be sensitive to randomness.
Meanwhile, the success set converges quickly: most recoverable cases succeed within two runs, and the total success set converges after three runs.
This suggests that a single additional repetition can substantially increase the overall success rate, while further repetitions provide diminishing returns.

%% file: sections/executioncontextmeasurement.tex
\section{Impact of Tooling and Input Signals}

In this section, we aim to answer \textbf{RQ2} by examining (1) how alternative designs (e.g., fewer tools for the agent) perform, and (2) the robustness of the solution to weaker signals from the input.

Understanding these effects serves two purposes.
First, it provides practical insights into which engineering components are most critical for building effective LLM-centric vulnerability reproduction systems.
Second, it allows us to assess whether LLM-based vulnerability reproduction can generalize to more constrained real-world environments, such as closed-source systems or restricted platforms where interactive debugging tools like GDB are unavailable.

To this end, we decompose the analysis into the following research questions:

\textbf{RQ2.1 (Tooling impact)}: How do reductions in agent capabilities affect effectiveness and trajectories? We ablate three components: (a) prompt quality (\S\ref{sec:rq23-badprompt}), (b) availability of subsystem-specific user-space utilities (\S\ref{sec:utility-removal}), and (c) interactive debugging support (\S\ref{sec:gdb-removal}).

\textbf{RQ2.2 (Input signal robustness)}: How does removing auxiliary patch context, i.e., commit messages, affect overall performance (\S\ref{sec:rq21-commitmsg})?

\subsection{Experiment Setup}

To study the impact of execution context on LLM-based vulnerability reproduction, we vary the prompt quality, availability of command line utilities, interactive debugging support, and patch commit messages.
For each variation, we conduct an independent set of experiments in which only the corresponding execution context is modified, while all other settings remain identical to those used in \S~\ref{sec:performeasure}.

\textbf{Prompt Quality.}
We use a much simpler prompt by retaining essential technical descriptions and execution constraints, while removing higher-level instructions.
Specifically, we omit instructions that prescribe an analysis workflow (e.g., performing static analysis before dynamic analysis), discourage premature assumptions, or encourage validation-driven reasoning.

\textbf{Subsystem-Specific Command Utilities Removal.}
We remove subsystem-specific user-space command-line utilities (e.g., \texttt{nft} for Netfilter and \texttt{tc} for traffic control) from the virtual machine disk image.
To ensure that the agent does not rely on such utilities, we explicitly instruct the agent not to use subsystem-specific command-line tools and block their invocation at execution time.

\textbf{GDB Removal.}
We remove the GDB toolset from the available execution environment and eliminate all GDB-related instructions from the prompt.
To avoid indirect use of debugging capabilities, we explicitly instruct the agent not to install or invoke GDB or other interactive debugging tools during the reproduction process.

\textbf{Commit Message Removal.}
We remove commit messages while retaining the corresponding code changes provided to the agent.
To prevent the agent from accessing commit messages through command-line tools (e.g., \texttt{git log}), we explicitly instruct the agent not to retrieve or infer commit messages during execution. 

For all these setups, we audit the recorded execution traces to verify that the agent complies with the specified instructions. We find that the agent consistently complies with our explicit constraints and does not attempt to circumvent them.

\subsection{Prompt Quality Degradation Evaluation}
\label{sec:rq23-badprompt}

Prior work~\cite{chainofthought} demonstrates that prompt design can significantly affect the effectiveness of agents. In the design of \systemname, we similarly provide high-level instructions that encourage the agent to perform more rigorous analysis, such as validating hypotheses and avoiding premature conclusions. To analyze their effects, we compare the baseline configuration against a degraded prompt that retains necessary technical content.

We find that prompt degradation leads to a modest reduction in success rates for both models. There is a 1 percentage-point drop for Medium (47.0\%$\rightarrow$46.0\%) and a 6 percentage-point drop for XHigh (56.0\%$\rightarrow$50.0\%). For Medium, degrading the prompt yields OR$=0.96$ with $p=0.31$, and for XHigh OR$=0.79$ with $p=0.17$. This suggests that the observed reductions in success rate are not statistically significant.

Despite the insignificant impact on overall performance, the tool invocation trajectory of the agent changes.  The number of source code browsing tool invocation decreases 6.5 times per case on average for Medium and 6.0 for XHigh.  We observe an even more pronounced shift in GDB breakpoint usage: under degraded prompts, breakpoint usage becomes rare in successful reproductions (Medium: 1 runs vs 38 in baseline; XHigh: 5 vs 33), while remaining relatively common in failed runs (Medium: 8 vs 22; XHigh: 16 vs 29).

According to the results, the overall usage of GDB drops substantially. Notably, unlike the baseline setting where GDB usage is more evenly distributed across outcomes, GDB becomes rare in successful reproductions and is concentrated in failed runs. This pattern suggests that, after removing prompts that encourage forming hypothesis and validation, GDB is no longer consistently used for verification and is instead primarily used in more difficult cases. 

Overall, results of the evaluation for \textbf{RQ2.1} shows that prompt degradation leaves aggregate success rates largely intact but makes the execution trace less hypothesis-driven: agents browse less code and postpone debugging until they encounter difficulty.

\subsection{User-Space Utilities Removal Evaluation}
\label{sec:utility-removal}

As suggested in Section~\ref{sec:tool-usage}, the agent commonly uses subsystem-specific user-space command-line utilities such as  \texttt{nft} and \texttt{tc} to interact with the VM. To examine this factor, we evaluate a restricted setting in which these utilities are removed and analyze the resulting impact on reproduction performance and the agent's behavior. 
We find that disabling subsystem-specific command utilities results in divergent but statistically insignificant changes in success rates. The success rate of XHigh decreases from 56.0\% to 46.0\% (OR$=0.67$, $p=0.20$) while Medium increases from 47.0\% to 50.0\% (OR$=1.13$, $p=0.78$). These results indicate that reproduction success may not depend on subsystem-specific command-line utilities, even though LLM agents tend to invoke them when available. 

Despite the overall effectiveness not being statistically affected by the removal of utilities, the form of PoCs changes substantially. Originally, in successful PoCs, command-line utilities are heavily used (31/56 for XHigh; 27/47 for Medium). After removing subsystem-specific utilities, invocation of command-line utilities almost disappears (2/46 for XHigh; 1/50 for Medium) from PoCs. In contrast, Table~\ref{tab:nocmdutil-header-shift} shows that successful PoCs more frequently include headers related to netlink, netfilter, and traffic control after utility removal. We find that the agents compensate for the absence of command-line utilities by directly leveraging syscalls and library APIs within the PoC.

\begin{table}[htbp]
\centering
\caption{Change in kernel-facing header usage in successful PoCs after removing subsystem-specific utilities.}
\label{tab:nocmdutil-header-shift}
\begin{tabular}{lccc}
\toprule
Model & Netlink & Netfilter & Traffic Control \\
\midrule
Medium & 37/50\,/\,15/47 & 14/50\,/\,6/47 & 20/50\,/\,3/47 \\
XHigh  & 34/46\,/\,21/56 & 17/46\,/\,9/56 & 17/46\,/\,6/56 \\
\bottomrule
\end{tabular}
\end{table}

We also observe that removing these utilities leads agents to inspect more source code while interacting less with the kernel. On average, source code browsing increases by 14.3 calls per case for Medium and 5.2 calls per case for XHigh. Meanwhile, interaction with the VM (including PoC uploading and command execution) decreases, with average reductions of 10.5 interactions per case for Medium and 20.7 for XHigh. 

We further inspect one representative case (commit \texttt{317eb968}) to understand agent trajectories. In the baseline setting, the agent makes an early mistake in its C-based PoC by constructing an incorrect \texttt{netlink} message, which fails to route requests to the intended subsystem. It then switches to the \texttt{nft} utility, bypassing the need to manually construct a correct \texttt{netlink} message, and still triggers the bug. When \texttt{nft} is unavailable, the agent can no longer rely on this workaround; instead, it inspects the kernel source code of the function \texttt{nfnetlink\_rcv\_batch}, identifies the dispatching logic, corrects the PoC, and reproduces the vulnerability using a pure C-based trigger. This case shows that agents adjust their reproduction strategies when certain tools are unavailable.

Overall, our evaluation shows that removing these utilities does not significantly affect aggregate reproduction success. Without these utilities, agents can still leverage syscalls and library APIs to craft PoCs.

\subsection{GDB Removal Evaluation}
\label{sec:gdb-removal}

GDB is commonly regarded as a critical tool for vulnerability analysis, motivating its inclusion in our agent toolset. However, whether agents can truly benefit from GDB during kernel vulnerability reproduction remains unclear. We therefore investigate the impact of completely removing GDB on both reproduction outcomes and agent behavior.

Through evaluation, the overall outcome is barely affected: the success rate of Medium does not change, and the success rate of XHigh drops by only one percentage point. This aligns with our observations in Section~\ref{sec:rq23-badprompt}, where agents use less GDB while overall performance remains intact.

To understand why GDB provides limited marginal benefit, we analyze agent trajectories and examine whether agents seek alternative sources of kernel visibility. We observe occasional attempts to access \texttt{kprobe}- and \texttt{ftrace}-related files, but these fail because such facilities are disabled; eBPF is only used for bugs directly involving it. Overall, we find no evidence that agents employ alternative mechanisms offering kernel-level visibility comparable to GDB.

Since kernel-level visibility is not a key factor to the performance, we further manually inspect 10 XHigh cases that succeed with GDB to understand how agents leverage them to facilitate PoC crafting.  We find that while agents occasionally encounter practical obstacles (e.g., variables reported as ``optimized out''), they can generally work around them and extract relevant kernel state. However, such inspections rarely lead to substantive corrections of the PoC, indicating that GDB serves to reaffirm already plausible hypotheses. We analyze the GDB usage and observe that GDB is primarily used to confirm PoC reachability or its influence on kernel values. However, such information rarely facilitates the reproduction of race conditions, where success is instead dominated by understanding and inducing the required concurrency.

Overall, the results suggest that while GDB is used to confirm reachability and inspect key kernel values, it is not a decisive factor for reproduction outcomes. Instead, success primarily depends on accurate root-cause analysis, and GDB rarely recovers trajectories once this understanding is incorrect. This explains why removing GDB has little impact on the outcome and why agents remain effective after GDB removal.

\subsection{Commit Message Removal Evaluation}
\label{sec:rq21-commitmsg}
Section~\ref{sec:vulfact} shows that informative commit messages significantly improve reproduction success. To assess agent performance in the absence of such information and evaluate the feasibility of applying LLM-based vulnerability reproduction for situations where commit messages are unavailable (e.g., Android OEM), we measure the impact of removing commit messages on reproduction outcomes.

After removal, the success rates of XHigh and Medium drop to 39.0\% and 36.0\%. Although the overall effectiveness remains reasonable, a notable degradation is observed. We further divide results through categories defined in Section~\ref{sec:vulfact} as shown in Table~\ref{tab:commitmsg-removal-4col}.

\begin{table}[ht]
\centering
\small
\caption{Effect of removing commit messages by disclosure type.
Each cell reports $\Delta$success in percentage points (without-commit minus with-commit),
followed by discordant-pair counts $(b/c)$ where $b$ is \emph{only-with-commit} successes and $c$ is \emph{only-without-commit} successes.
Type~1: $n{=}13$, Type~2: $n{=}79$, Type~3: $n{=}8$.}
\label{tab:commitmsg-removal-4col}
\begin{tabular}{lccc}
\toprule
 Model & Type~1 & Type~2 & Type~3 \\
\midrule
XHigh  & $0.0\%$ (1/1)   & $-19.0\%$ (19/4) & $-25.0\%$ (3/1) \\
Medium & $+7.7\%$ (0/1)  & $-12.7\%$ (16/6) & $-25.0\%$ (2/0) \\
\bottomrule
\end{tabular}
\end{table}

The results reveal that the impact of commit message removal is non-uniform across different levels of information disclosure. When commit messages provide little to no information about the vulnerability, their removal has limited effect on reproduction outcomes. In contrast, once messages contain trigger-related context, removing them leads to a significantly larger negative impact. Even without explicit reproduction guidance, the acknowledgment of a triggerable vulnerability and brief analytical context still contribute positively to successful vulnerability reproduction.

Overall, the results show that removing commit messages degrades reproduction performance, with larger drops observed when the original messages are more informative. Nevertheless, even without commit messages, the overall reproduction performance remains reasonably strong.

%% file: sections/beatsyzdirect.tex
\section{Comparative Evaluation}
To answer \textbf{RQ3}, we compare \systemname with SyzDirect, a state-of-the-art non-LLM approach for Linux kernel vulnerability reproduction.
Specifically, we contrast LLM-based and fuzzing-based approaches in terms of reproduction effectiveness and efficiency under aligned experimental settings, while accounting for their fundamental methodological differences.

\subsection{Measurement Setup}

Due to compatibility issues, the implementation of SyzDirect cannot be applied to newer kernels, making it difficult to evaluate newer vulnerabilities.

As shown in our cutoff-date analysis in Section~\ref{sec:performeasure}, the knowledge cutoff does not significantly affect reproduction outcomes.
Therefore, to enable a fair and practical comparison, we adopt the same Syzbot-based dataset used in the SyzDirect paper.
Following the Syzbot case identifiers reported in their evaluation, we crawled the corresponding vulnerability cases and rebuilt them under the kernel configuration and compiler settings described in Section~\ref{sec:measurementsettings}.

Due to the wide span of kernel versions covered by these cases, not all environments could be successfully reconstructed in our setup.
We filtered out cases whose corresponding virtual machine images could not be built or booted.
As a result, 85 cases remain from the original 100 cases.
We use these 85 cases as the denominator when computing reproduction success rates and compare \systemname’s results against the reproduction statistics reported in the SyzDirect paper.

\subsection{Performance Comparison}

We compare \systemname (XHigh) with SyzDirect in terms of reproduction effectiveness and efficiency to evaluate performance between LLM-centric and fuzzing-centric approaches in Linux kernel vulnerability reproduction.

For SyzDirect, we rely on the statistics reported in their paper and adhere to the evaluation protocol described therein.
Under this protocol, each vulnerability is executed for ten independent runs and is considered successfully reproduced if at least one run triggers the bug.
We use the reported mean time-to-exploit ($\mu$TTE) for successful cases, and assign a timeout of 24 hours to failed cases when computing the overall expected runtime.

Table~\ref{tab:comparison-performance} summarizes the comparison results.
On the 85-case aligned dataset, the LLM-centric approach, as instantiated by \systemname (XHigh), successfully reproduces 57 vulnerabilities, corresponding to a success rate of 67.1\%.
In contrast, the fuzzing-centric approach represented by SyzDirect reports 42 successful reproductions out of 100 cases in its original evaluation.

\begin{table}[htbp]
  \centering
  \small
  \setlength{\tabcolsep}{4pt}
  \begin{tabular}{cccc}
    \toprule
    Method & Success Rate & \makecell{Avg. Time \\ (Success)} & \makecell{Avg. Time \\ (Overall)} \\
    \midrule
    \systemname \\ (XHigh) & 57/85 (67.1\%) & 17.04 min & 22.48 min \\
    \midrule
    SyzDirect & 42/100 (42.0\%) & 11.74 h & 18.85 h \\
    \bottomrule
  \end{tabular}
  \caption{Performance comparison between \systemname (XHigh) and SyzDirect.}
  \label{tab:comparison-performance}
\end{table}

In terms of efficiency, LLM-centric approaches exhibit substantially lower reproduction time compared to fuzzing-centric approaches.
For successful cases, the LLM-based setup achieves an average reproduction time of 17.04 minutes, whereas the fuzzing-based setup requires an average of 11.74 hours.
When accounting for failed cases using a 24-hour timeout, the overall expected reproduction time is 22.48 minutes per case for the LLM-based approach, compared to 18.85 hours for the fuzzing-based approach.

We further observe that the LLM-centric setup maintains relatively low monetary cost.
Across all evaluated cases, the average cost per case is \$2.19, with successful reproductions costing \$1.58 on average.
Overall, these results characterize a fundamental efficiency gap between LLM-centric and fuzzing-centric vulnerability reproduction paradigms, while demonstrating that LLM-based approaches can achieve competitive effectiveness under practical evaluation settings.

\paragraph{Subsystems}

Notably, SyzDirect's dataset covers a more diverse set of kernel subsystems compared to the KernelCTF one, as Syzbot fuzzes the whole kernel without restricting the scope to interfaces accessible for non-privileged users.

We therefore aggregate reproduction success rates by subsystem and report those with more than three cases in Table~\ref{tab:xhigh-by-subsystem-3x4}.
\begin{table}[ht]
\small
\centering
\setlength{\tabcolsep}{2pt}
\begin{tabular}{|l|l|l|}
\hline
net 11/13 (84.6\%) & fs 5/8 (62.5\%) & mm 4/6 (66.7\%) \\
\hline
ext4 2/4 (50.0\%) & kernel 2/4 (50.0\%) & netfilter 3/4 (75.0\%) \\
\hline
wireless 4/4 (100.0\%) & block 3/3 (100.0\%) & btrfs 1/3 (33.3\%) \\
\hline
fbdev 2/3 (66.7\%) & ntfs3 1/3 (33.3\%) & squashfs 3/3 (100.0\%) \\
\hline
\end{tabular}
\caption{XHigh success by subsystem.}
\label{tab:xhigh-by-subsystem-3x4}
\end{table}
The results demonstrate that the agent remains effective even when applied to a broader diversity of subsystems.

\textbf{Success Rate} We observe that XHigh achieves a substantially higher success rate on the SyzDirect dataset than on the KernelCTF dataset.
A key reason is that the SyzDirect dataset is derived from Syzbot-reported vulnerabilities, which are vulnerabilities that Syzkaller can discover and generate bug reproducers. However, some complex vulnerabilities involving non-determinism, complex state accumulation behaviors are inherently hard to be reproduced as discussed by prior work~\cite{yome}. Consequently, such complex vulnerabilities are naturally excluded from Syzbot's dataset.

In contrast, many KernelCTF vulnerabilities involve complex race conditions and do not overlap with Syzbot-reported vulnerabilities. In our KernelCTF dataset, only three cases overlap with vulnerabilities that are discoverable by Syzbot. This discrepancy in dataset composition likely contributes to the substantially lower outcome on KernelCTF dataset.

Our results for \textbf{RQ3} show that LLM-centric vulnerability reproduction exhibits a distinct performance profile compared to fuzzing-centric approaches.
While both paradigms are capable of reproducing real-world Linux kernel vulnerabilities, LLM-based approaches achieve competitive effectiveness with substantially lower reproduction time and cost under practical evaluation settings.

%% file: sections/casestudies.tex
\section{Failure Analysis and Case Studies}
\label{sec:casestudies}

To further understand the reasons for agent failures, we conduct a manual investigation into the failed cases that never succeeded in any evaluated setting. To be practical, we limit ourselves to those cases where the corresponding KernelCTF technical write-ups exist. This leads to 13 cases in total. We find that 3 cases fail due to environment-related issues (e.g., kernel config differences), 2 fail because the agent cannot achieve the race condition within the short race window, and the remaining 8 fail due to incorrect or incomplete root-cause analysis. Among these categories, root-cause-analysis failures seems be the most important and prevalent one. 

To better understand root-cause-analysis failures, we conduct a deep dive into one representative case from the 13 consistently failed cases (Case Study I). We further present a second case that succeeds only under a single setting as a contrastive example. We use this case to highlight how subtle differences in tool usage during the root-cause-analysis phase can lead to divergent analyses and, consequently, different reproduction outcomes (Case Study II).

\paragraph{Case Study I: Failed Root-Cause Analysis from Misinterpretation.} 
\label{sec:casestudymisleaingcommit}

This case illustrates how agents' misinterpretation of the patch message and code diff causes the agents' root cause analysis to fail.
The commit of the patch is \texttt{e26d3009}.
Agents of all settings fail to trigger the bug.

The diff explicitly disallows a deactivation path for a struct \texttt{nft\_set} object. The patch message also mentioned ``anonymous sets'' (which are effectively struct \texttt{nft\_set} objects). 
From these signals, the agent infers that the vulnerability is a race targeting a struct \texttt{nft\_set} object, involving asynchronous garbage collection.
In particular, it hypothesizes that tearing down a set could free the struct \texttt{nft\_set} while GC work still holds a stale pointer, yielding a UAF on the \texttt{nft\_set} object.
The agent therefore focuses its effort on reproducing a UAF on struct \texttt{nft\_set}, but fails to trigger any corruption.

Our manual analysis indicates that the signals from the patch mislead the agent to believe the corruption target is struct \texttt{nft\_set}.
In fact, according to the actual KernelCTF exploit, the corruption is instead 
on a different but related object of type struct \texttt{nft\_chain}.
Unfortunately, since neither the patch message nor code diff mentioned \texttt{nft\_chain} explicitly, the agent failed to identify the correct object in question.

This suggests that the current solution is insufficient for reliable root-cause localization for complex vulnerabilities when the patch signals are low quality (or even misleading). To improve the solution against such cases, we believe the agents need hypothesis diversification, external stimuli (e.g., hints from static analysis), and early falsification loops that abandon overcommitted hypotheses when predicted corruption signals do not materialize.

\paragraph{Case Study II: Failed Root-Cause Analysis from Neglecting Source Code.}

This case illustrates how neglecting critical source code can cause agents' root cause analysis to fail.
The commit of the patch is \texttt{13114dc5}.
Interestingly, only XHigh (degraded prompt) successfully triggers the bug.

Triggering the bug requires the kernel to select a queue-based crypto backend (e.g., \texttt{cryptd}) that can reliably induce backlog behavior under high concurrency.
XHigh (degraded prompt) agent identifies this prerequisite because it inspects the relevant \texttt{crypto/} subsystem code, whereas other agents fail to do so due to shallow code investigation.

Specifically, only XHigh (degraded prompt) agent searches within the \texttt{crypto/} subsystem for backlog behavior, discovers that the behavior depends on which crypto implementation is selected, and therefore actively forces the use of a queue-based backend. This allows the generated PoC to trigger the bug.
In contrast, other agents primarily inspect interface-level definitions under \texttt{include/}. As a result, they do not recognize crypto backend selection as a key control point and do not attempt to steer the system toward \texttt{cryptd}, preventing the bug triggering.

This case indicates that future agent designs may benefit from stronger guidance toward analyzing the potential triggering chain implied by the patch. Such guidance can help agents reason toward concrete triggering conditions, instead of relying on randoms source code exploration that may miss key implementation-level details.

%% file: sections/related.tex
\section{Related Works}

\paragraph{LLM Agent for Security.}
Recent research increasingly integrates LLMs into autonomous agentic frameworks, enabling direct interaction with outside environments to various tasks. In the domain of offensive security, agents are designed to execute shell commands and analyze execution feedback, iteratively refining attack strategies against binary targets to generate exploits~\cite{yin2025pwngpt} or conduct penetration testing \cite{deng2025autopentester}. Parallel advancements in fuzzing demonstrate that agents can autonomously navigate complex configuration spaces and network protocols, utilizing documentation and error logs to guide state exploration where traditional evolutionary methods struggle~\cite{zhang2024prophetfuzz}. From the defensive side, it automates program repair and root cause analysis. Recent frameworks couple LLMs with debuggers and compiler toolchains; these agents interpret runtime signals and sanitizer reports to localize faults and verify patches through dynamic execution rather than static prediction~\cite{bouzenia2025repairagent, kim2025san2patch, yu2025patchagent}. These works demonstrate the effectiveness of LLM agents in performing cybersecurity tasks.

\paragraph{Automated PoC Generation}
Automated PoC generation aims to generate bug-triggering input that manifests a specific vulnerability. Traditional approaches primarily rely on fuzzing and symbolic execution. 

Fuzzing generates inputs, executes the program, and retains crashing inputs as PoCs~\cite{miller1990empirical}. Modern fuzzers often use runtime feedback: coverage-guided fuzzing improves exploration for bug finding~\cite{bohme2016coverage}, while directed fuzzing uses distance metrics to steer executions toward patch-related regions for patch validation and N-day reproduction~\cite{bohme2017directed}. Symbolic execution models inputs as symbolic variables, explores feasible paths, and uses SMT solving to satisfy path constraints and synthesize concrete triggering inputs~\cite{king1976symbolic,godefroid2005dart}. It is effective on highly constrained inputs~\cite{cadar2008klee,shoshitaishvili2016sok}, but often scales poorly. Hybrid systems combine fuzzing and symbolic execution to improve overall effectiveness~\cite{driller,symcc,qsym,mayhem}. More recently, AIxCC solutions show that LLMs can strengthen PoC pipelines by generating seeds and mutators and by assisting directed greybox fuzzing (DGF) target selection~\cite{kim2025atlantis,sheng2025all}.

%% file: sections/conclusion.tex
\section{Conclusion}

In this paper, we present the first large-scale measurement of LLM agents for Linux kernel N-day vulnerability reproduction using \systemname. Evaluating 100 real-world exploitable KernelCTF vulnerabilities, we find that agents successfully reproduce over 50\% of cases with practical time and monetary cost. Our analysis characterizes key failure modes, showing that race conditions and temporal memory bugs remain particularly challenging. Through controlled experiments, we observe that reproduction performance is largely robust to tool availability, whereas commit message information plays a critical role in guiding successful reproduction. Finally, on a SyzDirect-aligned subset, we demonstrate that LLM-based agents achieve higher reproduction rates with substantially shorter end-to-end time.